\documentclass[a4paper,11pt]{article}

\pdfoutput=1 

\usepackage{jinstpub} 
\usepackage{lineno}
\usepackage{url}
\usepackage{hyperref}
\usepackage{xcolor}
\usepackage{xspace}

\newcommand{\vtrxp}{VTRx+}
\newcommand{\mrad}{Mrad}
\newcommand{\fig}{Figure}
\newcommand{\mbps}{\ensuremath{\mathrm{Mb}/\mathrm{s}}\xspace}
\newcommand{\gbps}{\ensuremath{\mathrm{Gb}/\mathrm{s}}\xspace}
\newcommand{\micron}{\ensuremath{\upmu\mathrm{m}}\xspace}

\newcommand{\black}{\textcolor{black}}

\notoc

\title{\black{Characterization} of a high bandwidth readout chain for the CMS Phase-2 pixel upgrade}

\author{C. Smith}
\affiliation{\black{Department of Physics \& Astronomy, The University of Kansas, 1251 Wescoe Hall Dr., Lawrence, KS 66045, U.S.A.}}

\emailAdd{caleb.smith@ku.edu}

\abstract{
The CMS collaboration is building a new inner tracking pixel detector for the High-Luminosity LHC.
Each pixel readout chip will be controlled with a single serial input stream at 160~\black{\mbps} and will send out data via four current mode logic (CML) 1.28~\black{\gbps} outputs.
The readout chips will be grouped in modules and connected with up to 1.6~\black{meters} long low-mass electrical links to Low-Power Gigabit Transceivers (lpGBT) and Versatile Link PLUS Transceiver (\vtrxp) modules that send the data optically to off-detector electronics at 10~\black{\gbps}.
\black{The characterization of these components and system tests of the readout chain are presented.}
}


\arxivnumber{2110.14021} 

\collaboration[c]{\black{on behalf of the CMS tracker collaboration}}

\proceeding{Topical Workshop on Electronics for Particle Physics\\
  September 20--24, 2021\\
  Online event}

\begin{document}


\maketitle

\flushbottom

\newpage


\section{Introduction}
\label{sec:introduction}

In preparation for the High-Luminosity LHC (HL-LHC)~\cite{ref:hllhc}, the CMS Tracker will be fully replaced in the Phase-2 upgrade~\cite{ref:cms,ref:tdr,ref:orfanelli}.
The new inner tracker will have about two billion silicon pixels installed,
and hybrid pixel modules will process sensor data using a custom readout chip initially developed by the RD53 collaboration~\cite{ref:rd53}.
Each readout chip on the detector needs a control link to receive trigger, clock, commands, and settings from off-detector electronics as well as data links to send pixel data to off-detector electronics.
The high bandwidth control and data links are split into two stages: electrical links and optical links.
System tests of these electrical and optical links are presented.

\section{Data Readout Chain}
\label{sec:readout}

\begin{figure}[htbp]
\centering
\includegraphics[width=0.70\textwidth,origin=c]{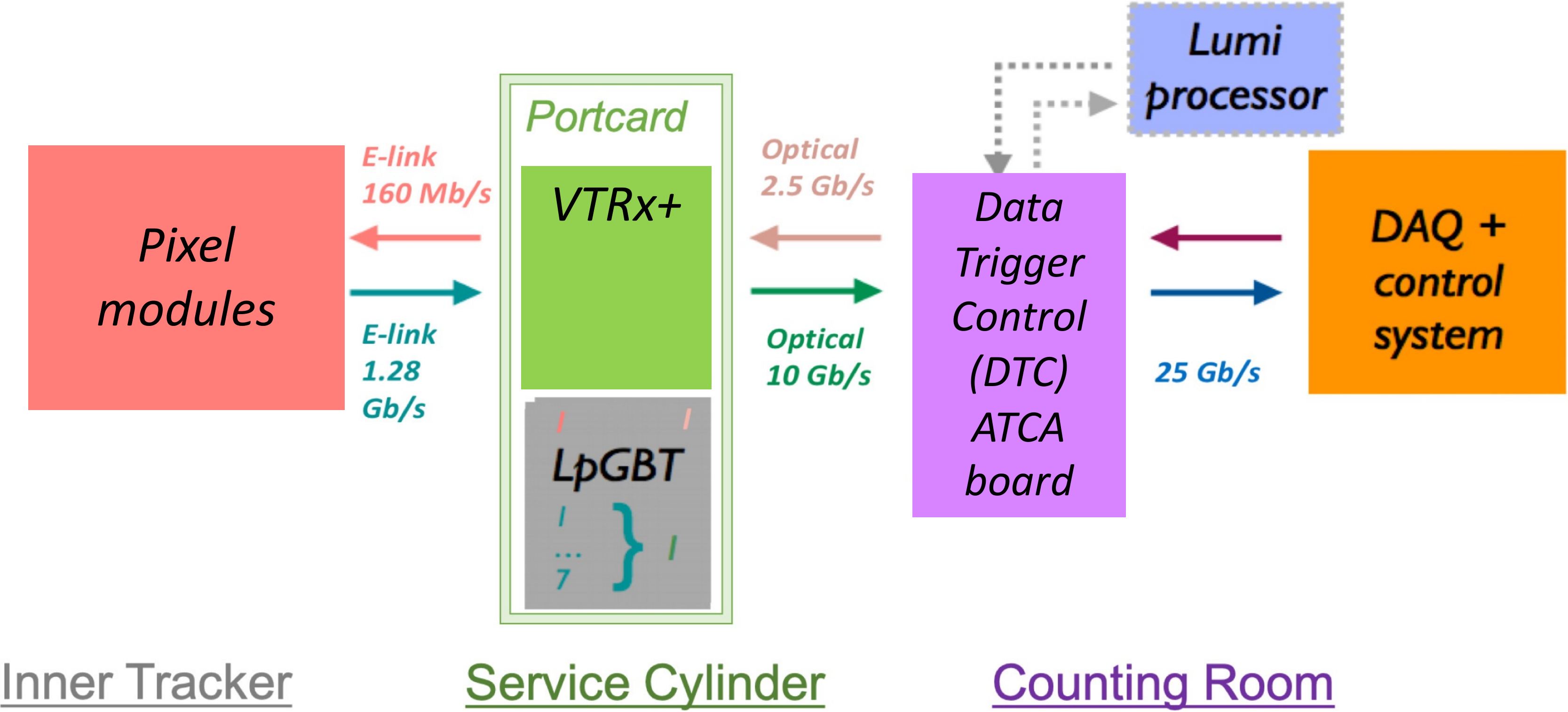}
\caption{
\label{fig:readout}
System readout architecture for the inner tracker~\cite{ref:orfanelli}. Pixel modules communicate with portcards through e-links. Portcards communicate with Data Trigger Control (DTC) boards through optical links.
}
\end{figure}

An overview of the data readout chain for the inner tracker is shown in \fig~\ref{fig:readout}.
High bandwidth electrical and optical links are used to establish control and data links between pixel modules on the CMS detector and Data Trigger Control (DTC) boards in the counting room.
Low-mass electronic links (e-links) up to 1.6~meters long provide 160~\black{\mbps} control links (downlinks) and 1.28~\black{\gbps} data links (uplinks) between pixel modules and portcards.
Optical fibers connect portcards, which are mounted on the support structure inside the detector, to DTC boards in the counting room to establish 2.5~\black{\gbps} control links (downlinks) and 10~\black{\gbps} data links (uplinks).
The portcards each carry three Low-Power Gigabit Transceivers (lpGBT)~\cite{ref:lpgbt_1} and three Versatile Link PLUS Transceiver (\vtrxp) modules~\cite{ref:vtrxp}.
\black{The lpGBTs communicate with the pixel modules via electrical links}, and the \vtrxp\space modules establish optical links with the DTC boards.

\section{Electrical Links}
\label{sec:electrical}

\begin{figure}[htbp]
\centering
\includegraphics[width=0.4\textwidth,origin=c]{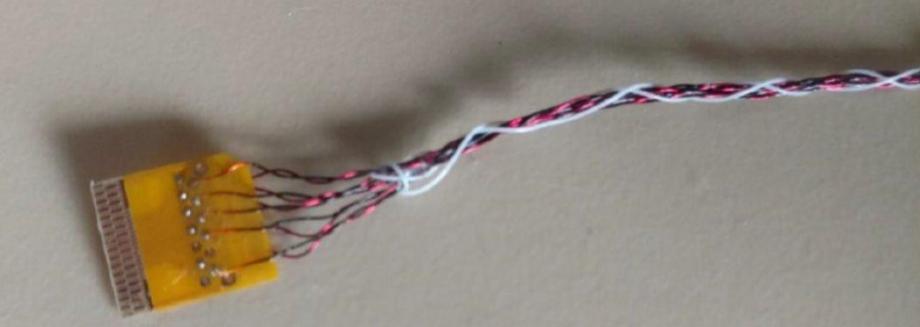}
\caption{
\label{fig:elink}
\black{One end of an electrical link (e-link) bundle that has five twisted pair channels: one channel for a control link and four channels for data readout.}
}
\end{figure}

The electrical link (e-link) \black{bundles} are created using low mass, small diameter twisted pair cables.
One end of a twisted pair electrical link \black{bundle} is shown in \fig~\ref{fig:elink}.
E-link bundles are made to connect to modules and provide each module with one control link twisted pair and multiple data link twisted pairs.
Each twisted pair is produced using two 36 American Wire Gauge (AWG) copper core (127~\black{\micron} in diameter) wires surrounded by polyimide insulation (45~\black{\micron} thick) that are twisted together with 4 twists per inch.
The pairs are soldered to small, 20~\black{\micron} thick PCBs that can plug into Molex connectors with 300~\black{\micron} pitch~\cite{ref:molex45}.
To improve durability and handling, the soldered wires are secured with a non-conductive epoxy, and then the bundle is lashed together using polyimide braiding.

Prototype low-mass electrical links \black{are characterized using} a series of measurements.
The signal quality is assessed using eye diagrams, which overlay transitions between binary states from repeated measurements of the signal over a time interval.
Bit error rate scans are used to characterize signal integrity across e-links.
\black{To reliably readout pixel detector data, the e-links need to support error rates of less than $10^{-13}$ for signals at 1.28~\gbps.}
Cross talk effects from channels within an e-link bundle (internal) and from multiple bundles (external) are studied.

\begin{figure}[htbp]
\centering
\includegraphics[width=0.44\textwidth,origin=c]{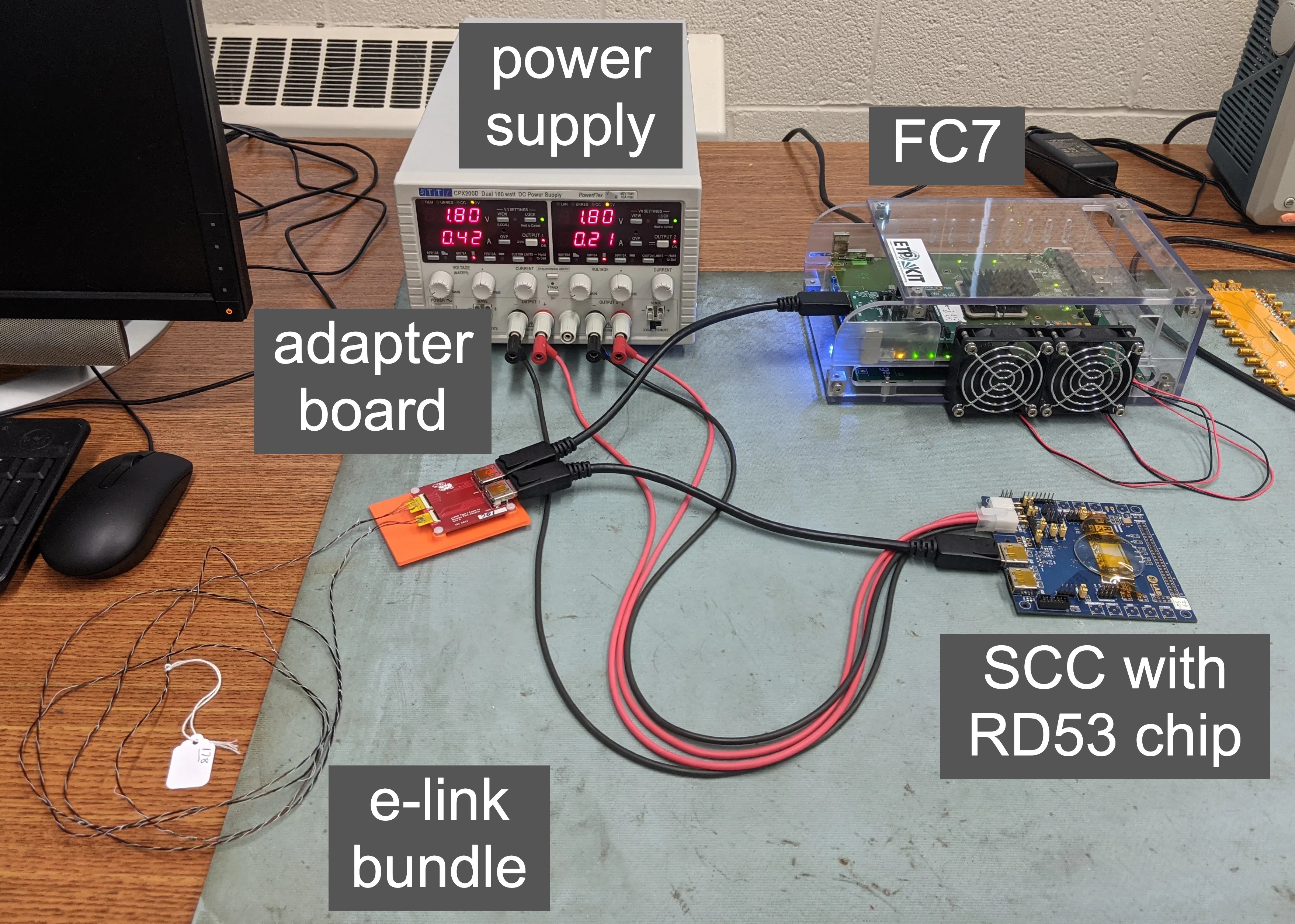}
\qquad
\includegraphics[width=0.32\textwidth,origin=c]{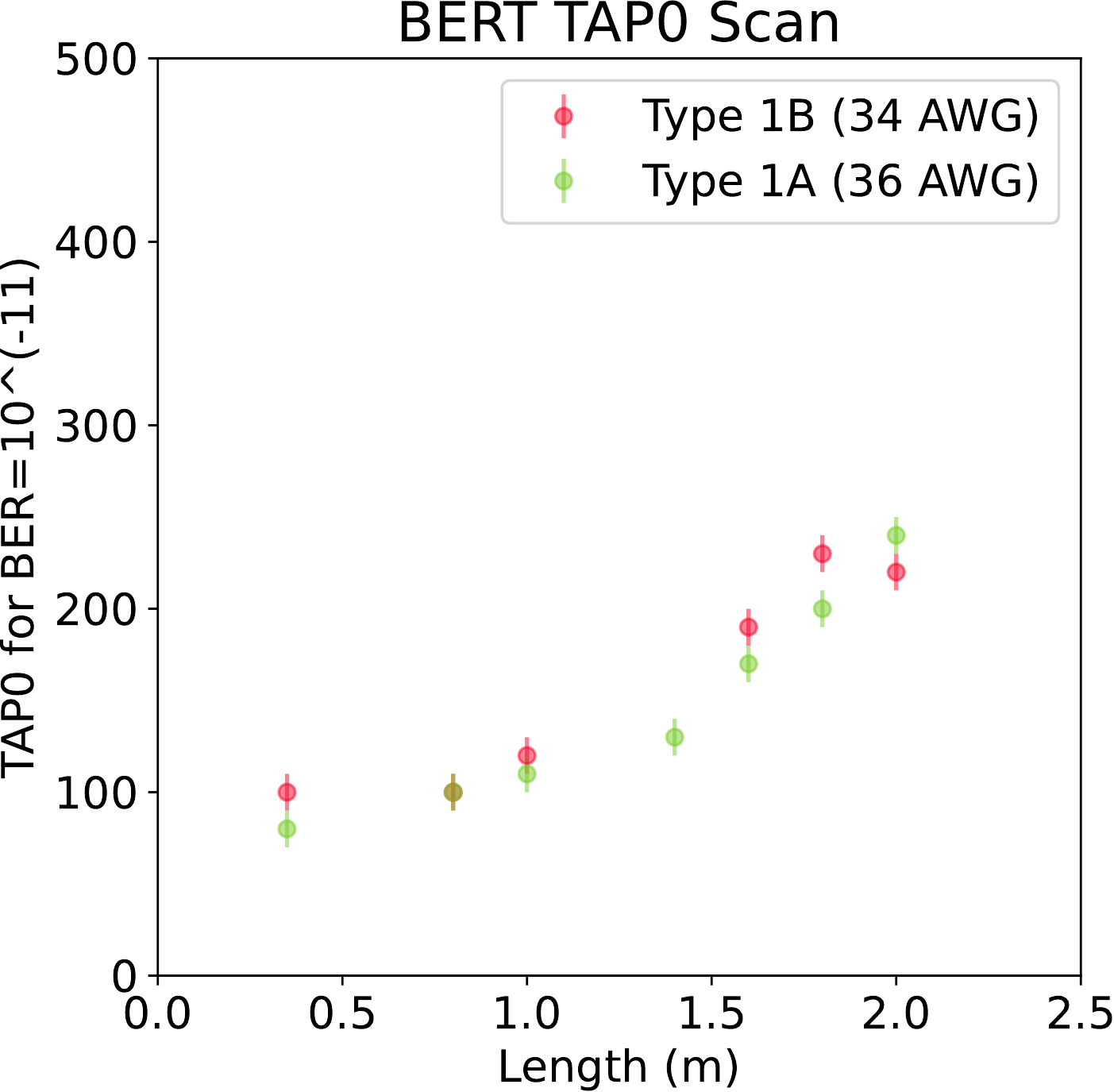}
\caption{
\label{fig:tap0_vs_length}
Measurement of signal amplitudes (set by TAP0) on e-links for a fixed bit error rate.
In the hardware setup (left), an FC7 mezzanine card is connected to an SCC through two commercial display port cables, an adapter board, and an e-link \black{bundle} for control and readout of an RD53 chip.
The measurements (right) show the TAP0 setting to achieve a bit error rate of $10^{-11}$ for e-links of different lengths and gauges.
}
\end{figure}

\black{To test error rates over e-links}, an RD53 chip on a Single Chip Card (SCC) is read out over e-links using an FC7 mezzanine card~\cite{ref:fc7}, as shown in \fig~\ref{fig:tap0_vs_length}.
E-links of two gauges (34 and 36~AWG) and different lengths (from 0.35 to 2.0~meters) are used in the readout chain.
Bit error rates are determined using a pseudorandom binary sequence (PRBS) that is sent \black{at 1.28~\gbps} from the RD53 chip to the FC7 mezzanine card.
The amplitude of the PRBS signal is varied using the \black{10-bit ``TAP0'' pre-emphasis setting}, which controls the amplitude of the signal output by the current mode logic (CML) driver on the RD53 chip.
For this test, the amplitude is increased until the bit error rate of $10^{-11}$ is reached.
Longer e-links require a larger signal amplitude to maintain a given bit error rate.
Good performance is seen for e-links up to 2.0~meters.
\black{A different hardware setup with a KC705 has been used to demonstrate bit error rates of less than $10^{-13}$, but longer tests using the RD53 chip are needed to reach this same target.}

\begin{figure}[htbp]
\centering
\includegraphics[height=2.0in,origin=c]{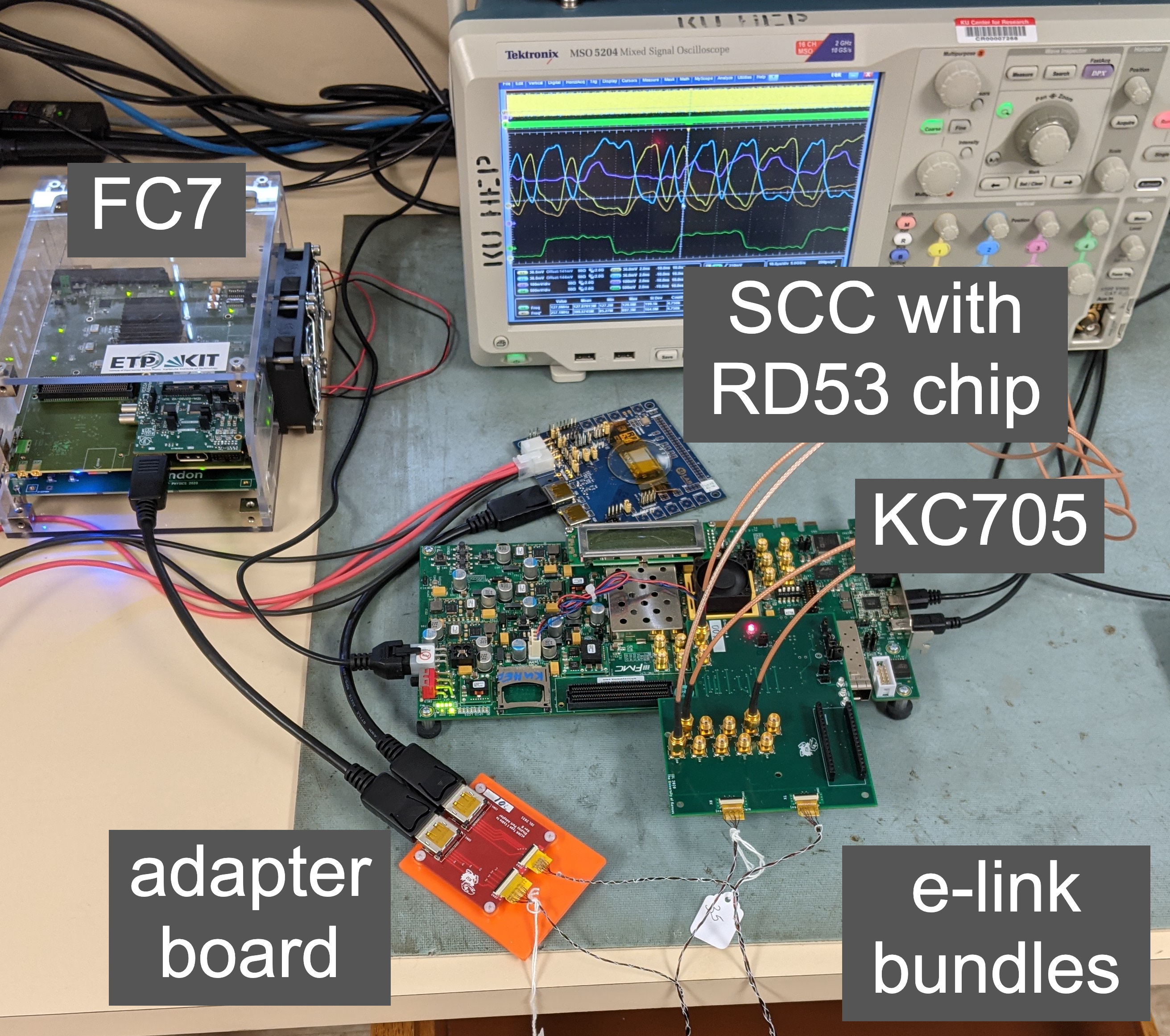}
\hspace*{.2in}
\includegraphics[height=2.0in,origin=c]{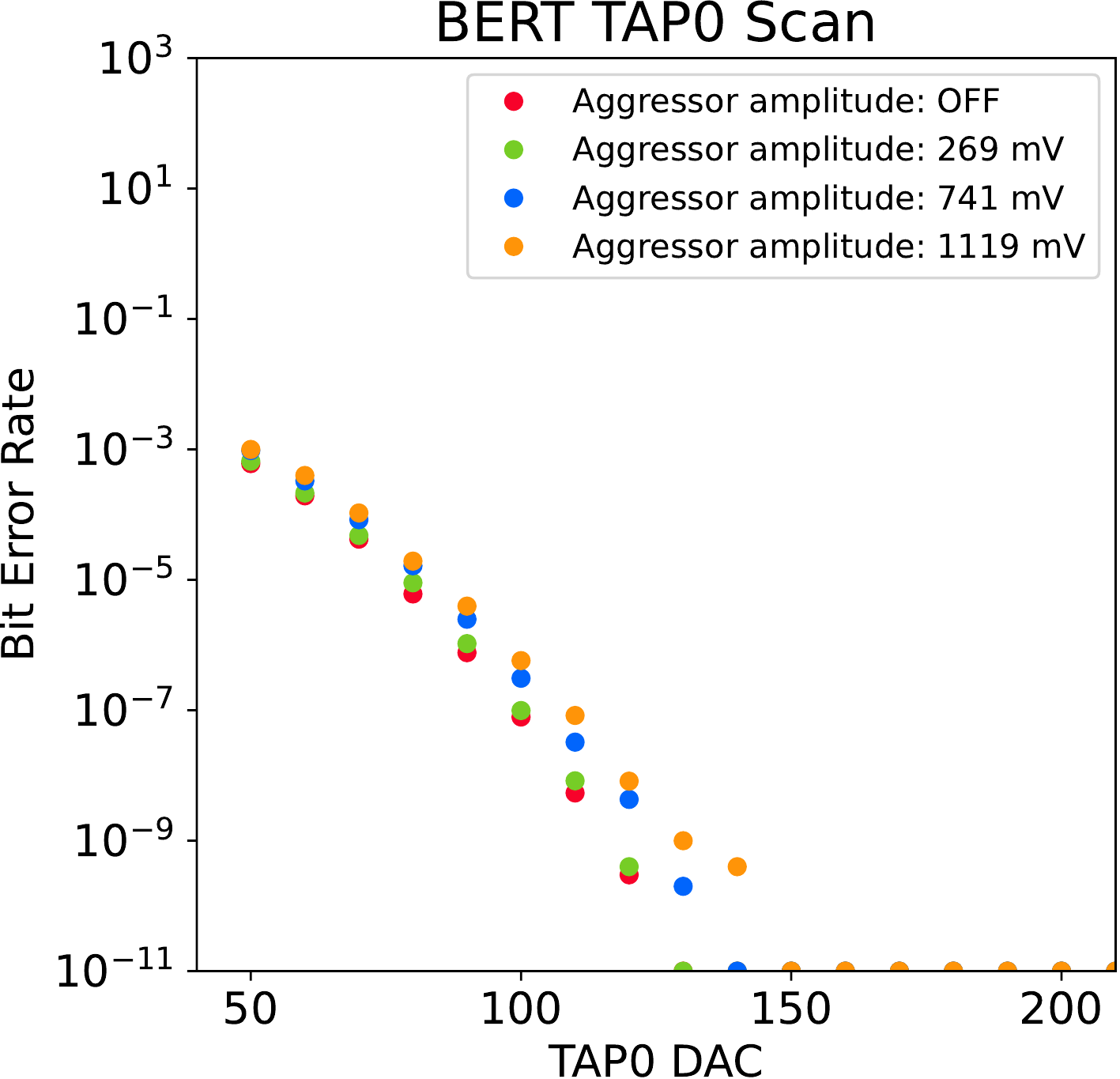}
\caption{
\label{fig:external_crosstalk}
Measurement of the effect of external crosstalk on data transmission across e-link \black{bundles}.
Two e-link \black{bundles} (1.4~meters, 36~AWG) are twisted together (left), with a victim e-link \black{bundle} connected between an FC7 mezzanine card and SCC to read from an RD53 chip, and an aggressor e-link \black{bundle} connected to a KC705.
The KC705 sends PRBS signals on four e-link channels on the aggressor at different amplitudes.
}
\end{figure}

Furthermore, an external crosstalk measurement for \black{e-link bundles} was performed, and the results are shown in \fig~\ref{fig:external_crosstalk}.
To mitigate the effect of external crosstalk from a large aggressor amplitude, the victim amplitude set by TAP0 only requires a small increase.
Thus, external crosstalk from the single aggressor e-link \black{bundle} has a small effect on readout over the victim e-link \black{bundle}.

Radiation testing was performed to determine if e-links can maintain performance \black{up to the design specification of} 1500~\mrad.
Two different epoxies, Araldite 2011~(Ref.~\cite{ref:araldite}) and UR6060~(Ref.~\cite{ref:ur}), were used for e-links that underwent radiation testing.
\black{After} 1300~\mrad, the Araldite 2011 epoxy showed significant darkening, while the UR6060 epoxy remained transparent.
The signal integrity on e-links remained good after \black{radiation}.
The polyimide braiding used for lashing becomes brittle after 1900~\mrad, but this should not \black{affect} the electrical readout.

\section{\black{Portcards}}
\label{sec:optical}

\begin{figure}[htbp]
\centering
\includegraphics[width=0.70\textwidth,origin=c]{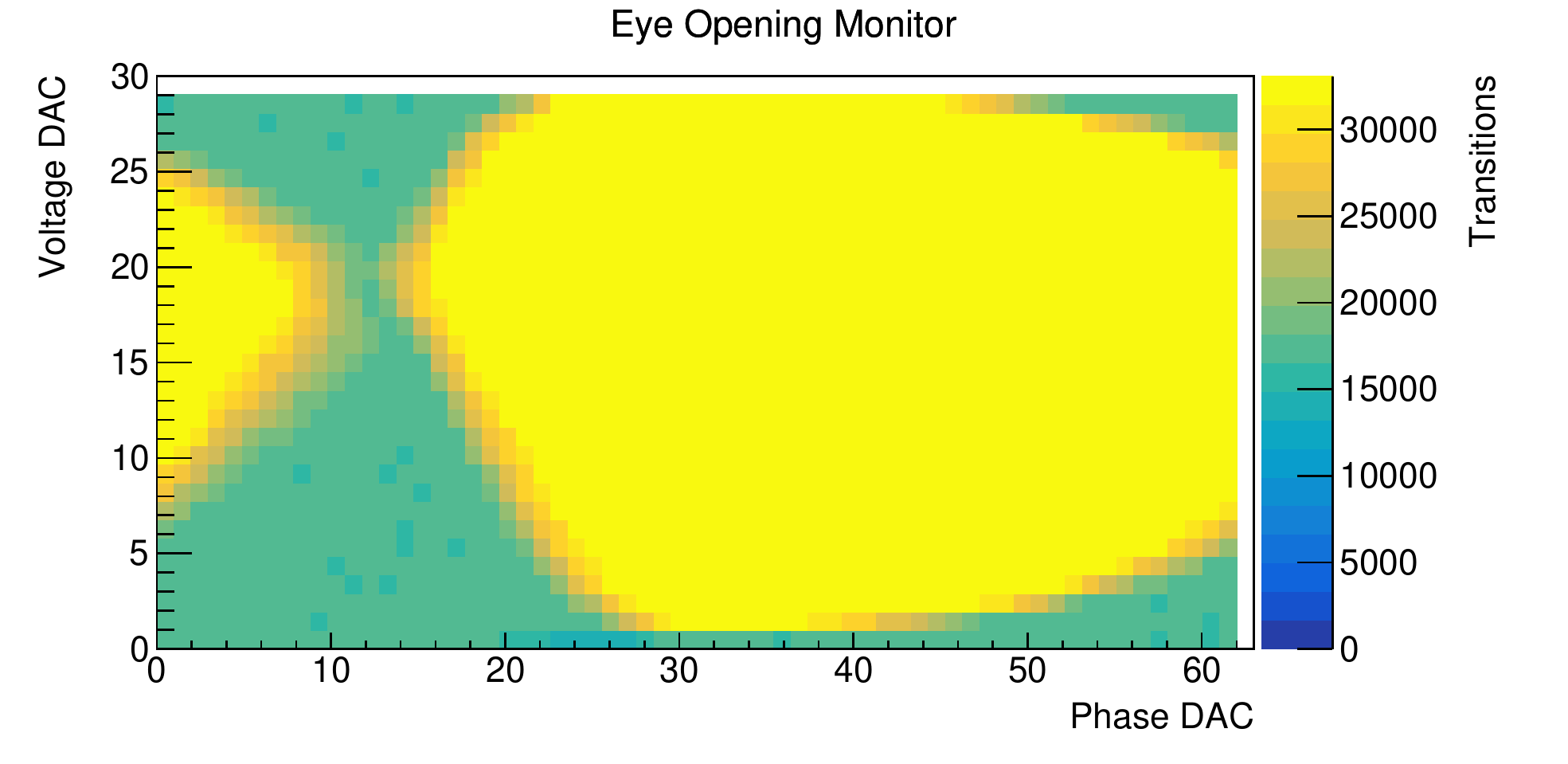}
\caption{
\label{fig:lpgbt_eye}
Eye opening monitor from an lpGBT for the 2.56~\black{\gbps} optical down link with equalization parameters at default settings.
The vertical axis is an encoded voltage threshold, and the horizontal axis is an encoded sampling phase.
The color scale shows the number of transitions (toggle count).
The plot is scaled vertically to show the eye opening region, and the regions without transitions are not shown.
The regions with high toggle count correspond to the eye opening.
}
\end{figure}

The portcards convert electrical signals from on-detector readout chips to optical signals for off-detector DTC boards (see \fig~\ref{fig:readout}).
Dedicated setups with the portcards reading out SCCs with short coaxial cables have been used to test the prototype portcards and validate the optical part of the readout chain.
A measurement of the optical links using the lpGBT eye opening monitor~\cite{ref:lpgbt_1} was performed in which the high-speed input to the lpGBT is compared to an adjustable threshold voltage by a comparator sampled with a phase interpolated clock.
Transitions of the comparator are counted, giving a ``toggle count.''
Within the eye, a large number of transitions are expected, and outside of the eye, approximately half as many transitions are expected.
Data from this measurement are shown in \fig~\ref{fig:lpgbt_eye}.
The vertical axis is scaled such that regions with no transitions are not shown.
The high toggle count region is large and corresponds to a large eye diagram and a strong optical signal.

%


\begin{figure}[htbp]
\centering
\includegraphics[width=1.0\textwidth,origin=c]{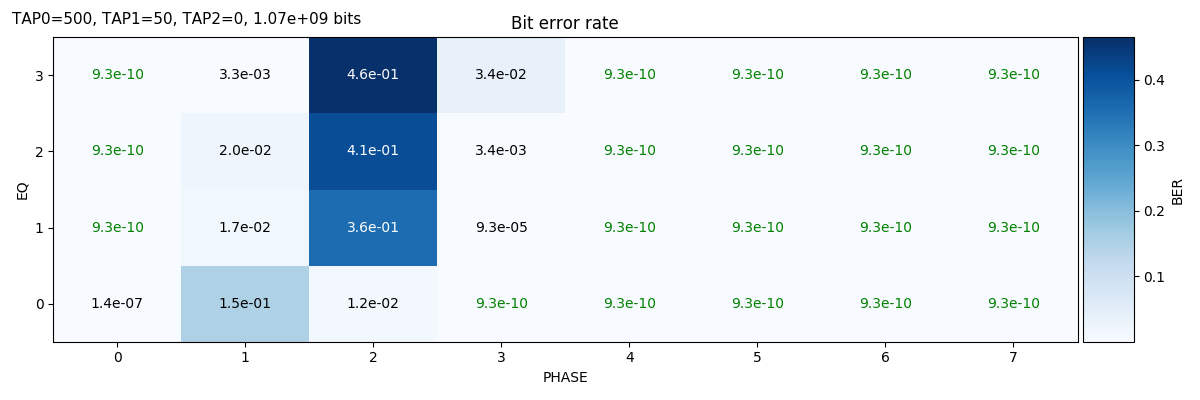}
\caption{
\label{fig:lpgbt_bert}
Scan of \black{bit error} rate vs. clock sampling phase and equalization setting for electrical port receivers on the lpGBT driven by the RD53 pixel readout chip.
The number of bits checked at each setting is $1.07 * 10^{9}$.
Points in green have zero observed errors and report $9.3 * 10^{-10}$, which is the reciprocal of the number of bits checked.
}
\end{figure}

In addition, the electrical link receivers on the lpGBT are being studied.
In one measurement, the signal source is a PRBS7 generator in the RD53 readout chip.
This signal is sent over a commercial display port cable, an adapter board, and a \black{commercial} Molex FPC cable (Ref.~\cite{ref:molex_cable}) to be received by an lpGBT.
The signal is compared to an internal error checker on the lpGBT.
The \black{bit error} rate is measured as a function of clock sampling phase and equalization setting, and the results are provided in \fig~\ref{fig:lpgbt_bert}.
The region with a bit error rate of $9.3 * 10^{-10}$ corresponds to zero errors for the total number of bits checked ($1.07 * 10^{9}$).
Low error rates on the lpGBT electrical link receivers are achieved by tuning the sampling phase, and changing the equalization setting has only a small effect on the timing.

\section{Conclusion}
\label{sec:conclusion}

Various system tests have been performed \black{for the new pixel readout chain.}
Electrical links provide stable connectivity at the required lengths, and crosstalk has a small effect on signal quality.
\black{Low bit error rates ($10^{-11}$) over e-links are observed, but longer tests with the RD53 chip are needed to achieve the target error rate ($10^{-13}$).}
Conversion from electrical to optical links is working well, and the optical links have good performance.
These results form an important input to the final design, production, and testing of components for the pixel detector readout chain.

\acknowledgments

The authors would like to thank the National Science Foundation for funding this research.


\bibliographystyle{JHEP}
\bibliography{twepp_paper}

\end{document}